# More room at the top: how small buoys aim at the detailed dynamics of the air-sea interface

Luigi Cavaleri[a]*, Victor Alari[b], Alvise Benetazzo[a] Jan-Victor Bjorkqvist[c], Oyvind Breivik[c,d], Jacob Davis[e], Gaute Hope[c], Atle Kleven[f], Frode Leirvik[f],Tor Nordam[f,g], Jean Rabault[h], E.J. Rainville[e], Sander Rikka[b], Torunn Irene Seldal[c], Jim Thomson[e]

a – CNR-Institute of Marine Sciences, Venice, Italy

b – Dpt. of Marine Systems, Tallinn Univ. of Technology., Tallinn, Estonia

c – Norwegian Meteorological Institute, Bergen, Norway

d – Univ. of Bergen, Bergen, Norway

e – Applied Physics Lab., Univ. of Washington, Seattle, Washington, USA

f – SINTEF Ocean, Trondheim, Norway

g – Dpt.t of Physics, NTNU, Trondheim, Norway

h – Norwegian Meteorological Institute, Oslo, Norway

* corresponding author: luigi.cavaleri@ismar.cnr.it

August 29, 2024

**Abstract**

Air-sea exchange processes have been identified as essential for both short- and long-term atmospheric and ocean forecasts. The two phases of the fluid layer covering our planet interact across a vast range of scales that we need to explore to achieve a better understanding of the exchange processes. While satellites provide a distributed large-scale view of the sea surface situation, highly detailed measurements, e.g., from oceanographic towers, are necessarily local. The required intermediate solution (i.e., data that are both accurate and distributed) can be provided by swarms of miniature surface buoys. As size, weight, and cost are reduced, these can be deployed in large numbers to investigate specific processes that are at present only crudely parameterized in our models as a result of scarcity of good measurements. Perhaps the most crucial process is white-capping in stormy conditions, where air-sea exchanges are enhanced by one or two orders of magnitude. Other applications include wave-current interactions, wave-ice interactions, and plunging breakers in the coastal zone.

Following a dedicated workshop, we summarize here the main findings and possibilities derived from the different approaches, and summarize the state of the art for a selection of miniature buoys. We list the solutions presented at the workshop, as well as other similar buoys, with their main characteristics and range of application. We describe the various possibilities of practical use and the scientific and engineering problems to be solved. Looking to the future, we also point out where the present technological improvements are leading to.

**Significance statement**

The interactions between ocean and atmosphere, with the continuous and intense exchanges of energy, humidity, air and spray, are fundamental in controlling the Earth climate. Their quantification is the basic information for any long-term forecast. Satellites provide large-scale views of the situation, but we need detailed local measurements at the sea surface for a full picture of the situation. Some isolated oceanographic towers do provide detailed pictures of their local conditions, but distributed high-density information is also needed. The present best solution is provided by miniaturized instrumented telemetering buoys, cheap and potentially expendable, to be distributed in large numbers (tens of them or more) where a specific forecast (e.g., hurricanes) or a scientific purpose (e.g., wave breaking in a stormy sea) suggest to act. While the present order of magnitude of size of these buoys is around 10×10×10 cm$^3$, the continuous technological improvement suggests that in the near future volumes could be one or two orders of magnitude smaller, opening for new problems and possibilities.

1. **The need for more measurements at the sea surface**

Air-sea exchange processes are widely recognized as essential for both short- and long-term atmospheric and ocean forecasts. This holds for processes ranging from the local generation of wind waves to the development of a low-pressure atmospheric system or of a tropical storm, and all the way to planetary scale processes such as the Gulf Stream or El Niño. Whichever the scale, all these processes depend on the exchange of thermal and mechanical energy, momentum, and mass between the two fluid layers that envelop our planet.

Such recognition has not always been the case. In the 1970s, Erik Mollo-Christensen (MIT) quipped that "for the oceanographers the atmosphere is a place where wind blows—for the meteorologists the ocean is a wet surface" (Cavaleri et al. 2012). In the realm of parameterizations, a turning point came around 1990 when the European Centre for Medium-Range Weather Forecasts (ECMWF) showed that coupling the wave and atmospheric models led to improved weather forecasts (Janssen 1991). Of course, we have come a long way since then. Presently satellites monitor the ocean surface daily, providing extensive data for the initialization of coupled atmospheric and ocean medium-range and seasonal forecasts. However, the problem is still to find a sufficiently detailed description of the physics at the interface, most of all how the wind and the waves it generates interact and give rise to all the exchanges that control the evolution of both the atmosphere and the ocean.

Possibly the strongest limits to the accuracy of the long- and short-term forecasts is given by the accuracy with which we are able to define the surface exchanges (Magnusson et al. 2019). Of course, more or less sophisticated parameterizations exist, based for a long time mostly on wind speed (e.g., Ma et al. 2017, Lee et al. 2022, and Troitskaya et al. 2023), recently also considering the impact of the sea state (Breivik et al. 2015, see also Edson et al. 2013, and the COARE 3.6 algorithm at https://github.com/NOAA-PSL/COARE-algorithm, as well as Pineau-Guillou et al. 2018). These are however still parameterizations that suffer from a lack of detailed observations. Although some oceanographic towers and platforms are equipped with highly detailed instruments to measure air, water and surface characteristics (see, among others, the Acqua Alta tower in the Gulf of Venice (Cavaleri 2000; Cavaleri et al. 2021), and Ekofisk in the central North Sea (Malila et al. 2022b, 2023), these and other long-term oceanographic stations such as Ocean Station Papa in the Northeastern Pacific (Thomson et al. 2013; Schwendeman and Thomson 2014, 2017) are too few and limited in their ability to yield accurate measurements of the wave field under sufficiently varied conditions.

The crucial point is that it is not yet clear at which resolution we need to resolve the interaction of wind and waves in the open ocean. Wave breaking entrains air bubbles at scales of millimeters and smaller. Laboratory experiments have clearly shown (Troitskaya et al. 2017) how we need to go down to the micrometer scale to describe the spray production in wind wave generation. These are not problems we can fully solve tomorrow, but they show the direction we need to go if we want to improve our physical description and parameterizations of the processes at the interface.

As mentioned above, a large part of the present data at the interface is acquired by buoys, from the largest specimens such as the Floating Instrument Platform (FLIP, now decommissioned; see Lenain and Melville 2017) or the W1M3A multidisciplinary observatory (Cavaleri 1984, www.w1m3a.cnr.it/OI1/modules/site_pages/about.php), to the smaller ones such as the Waverider (see Joosten 2013 for a historical account of the Waverider buoy) that at the time revolutionized wave data collection.

The technological evolution has since led from the once classical size, order of 1 m and a few hundred kg, to the smaller and lighter Spotter buoy (Raghukumar et al. 2019, http://www.sofarocean.com). These are presently deployed in large numbers in the world oceans, providing measurements for data assimilation and forecast validation (Houghton et al. 2021, 2022). It is interesting to note that 10,000 of these buoys would, for an overall cost of USD 50 M, provide much of, if not more than, the information provided by a typical satellite which costs ten times as much. Such buoys could not replace satellites, but the cost of such a complementary deployment is now well within reach of a large-scale scientific program.

However, these smaller buoys have only been an intermediate step. Technology moves ahead, and it is now possible to make most of, if not all, these measurements with even smaller, lighter, and hence cheaper buoys. Following a dedicated meeting on Wave Buoys and Open Source Platforms (WOOP24) in Finse, Norway, in February 2024, the purpose of this paper is to describe some of the presently available solutions and to point out the scientific opportunities that come with these miniature buoys.

Small buoys are largely made possible by the collapse in price (thanks to mass-market consumer-grade products using similar chips) and ease of use (thanks to open source and "makers" communities) of global navigation satellite system (GNSS) units, inertial motion units (IMU), microcontroller units (MCUs), and electronics components in general. The same trend is found in, e.g., thermistors, conductivity (salinity), hygrometry, and wind sensors. This allows us to build affordable buoys collecting a full range of data and parameters.

Physical oceanography has benefited enormously from new technology over the past century. As Walter Munk said in the documentary "One Man's Noise: Stories of An Adventuresome Oceanographer" (1994, https://www.youtube.com/watch?v=je3QvqNdHl0), "most of the progress that's been done in oceanography was the result of applying some new technology, not because somebody had some great new ideas." He went on saying "A new technology almost always leads to new understanding". We believe that by making our buoys smaller, lighter, cheaper, and open source, progress will be made by more people having access to more observations at ever smaller scales.

**2. The present technological solutions**

We list and briefly describe here the main characteristics of the buoys presented at WOOP24. The purpose is to provide the background for the related experiments and data described in the two following sections. Figure 1 shows the five buoys whose details are given in Table 1.

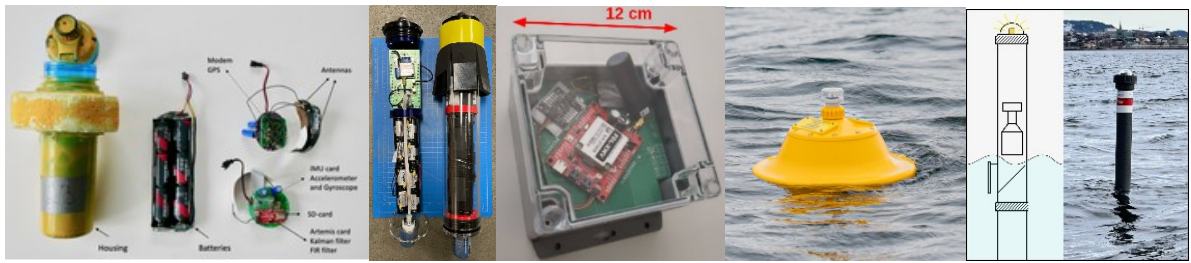

Figure 1 – The five buoys whose performance and characteristics have been discussed at the Finse meeting. Their main characteristics are detailed in Table 2. From left to right: SFY, microSWIFT, OpenMetBuoy (OMB), LainePoiss, and Spartacus.

Table 1. Main characteristics of the five described buoys, as shown in Fig. 1. Acronyms definition: pos.: position measurement; temp.: temperature measurement; TCO: total cost of ownership; FOSSH: free open source software and hardware; FOSS: free open source software.

|  | SFY | MicroSWIFT | OMB | LainePoiss | Spartacus |
|---|---|---|---|---|---|
| Size | 8x8x30 cm | 9x9x50 cm | 10x10x12 cm | 36x36x30 cm | 11x11x600 cm |
| Weight | 0.5 kg | 2.9 kg | 0.5-1 kg (batt. choice) | 5.5 kg | ~30 kg |
| Parameters measured | acceleration time series & pos | 2D spec., SST, SSS | 1D spec, pos., temp. | 2D spec., acceleration | particles and bubbles, buoy motion and position |
| Sampling frequency | 208 Hz | 4 Hz | 800 Hz raw, 10 Hz spectrum | 50 Hz | configurable, up to 20 Hz |
| Communication methods | GSM | Iridium | Iridium | GSM, Iridium | GSM |
| Power source | D-cell, Li. or Alk. | C-cell | D-cell, Li or Alk. | rechargeable batteries | D-cell, Li or Alk. |
| Autonomy |  | 2 months | up to 1+ year | >8 months | weeks or months |
| Cost estimate (USD) | 250 | 1000 | 650 hardware, 1000 TCO | 5500 | 5000 |
| Licence | FOSSH | FOSS | FOSSH | MIT licence for processing scripps | under development |

a. SFY

The Small Friendly Buoy (SFY, Hope et al. 2024) is a lightweight wave buoy (∼ 0.5 kg) aimed at coastal deployments within coverage of the cellular network. The buoy measures waves using an IMU at 208 Hz, and can transmit back the time series of the sea-surface acceleration at 52 Hz. The very high frequency, and the low cost due to using the cellular network, enables the buoy to measure the waves and wave field in a phase-resolved way. The individual phases and breaking events can be tracked across buoys, and the high frequency makes it possible to study the trajectory of a buoy also through plunging waves at beaches. In total, 48 SFYs have been built so far, tested in Europe, Africa, Asia, on Hawaii in the Pacific Ocean, and in the North Sea with connectivity from oil platforms. All code and processing scripts, including hardware schematics, are available at https://github.com/gauteh/sfy. The water-following capability of the buoy has allowed it to measure accelerations in excess of $10g$ within plunging breakers, (Hope et al. 2024).

b. microSWIFT

The microSWIFT buoy is a miniaturization of the Surface Wave Instrument Float with Tracking (SWIFT) first introduced in Thomson (2012) and revised in Thomson et al. (2019). The microSWIFT is described in detail within Thomson et al. (2023). The primary features are GNSS-based directional 2D wave spectra and modular additional sensor payloads (presently water temperature and conductivity, from which salinity is calculated). Intended future payloads include light and turbidity sensors. The form-factor of the microSWIFT is specifically designed for aerial deployments; the cylindrical shape fits the dropsonde tube of most research aircrafts. To date, 110 microSWIFTs have been built and deployed world-wide. A repository of quality-controlled data is maintained at https://doi.org/10.5061/dryad.jdfn2z3j1 and all code is public at https://github.com/SASlabgroup.

c. OpenMetBuoy

The OpenMetBuoy (OMB, see Rabault et al. 2022) is a small (typically 10 × 10 × 12 cm), lightweight (around 0.5 to 1 kg), low-cost, low-power (3 Li D-cells last more than 6 months), Iridium-SBD-connected, open software and hardware (https://github.com/jerabaul29/ OpenMetBuoy-v2021a) buoy, developed from previous waves-in-ice instruments (Rabault et al. 2016, 2017, 2020). Since sea ice typically blocks the horizontal wave orbital motion, OMB uses vertical IMU acceleration to estimate the 1D vertical wave spectrum. Particular care is taken to resolve small motion in calm conditions. Waves typically under 0.5 cm amplitude at 16 s period have been resolved (Rabault et al. 2022). The OMB is also able to accurately resolve 1D spectra of open ocean waves in all conditions.

The OMB design is fully open and it has for example been modified to log sets of several temperature sensors (Muller et al. 2024). Fully assembled OMBs can also be purchased from a commercial company (https://www.labmaker.org/products/openmetbuoy – no commercial ties to the authors).

d. LainePoiss

The development of LainePoiss (Alari et al. 2022) wave buoys started in early 2018 aiming at a robust instrument to measure waves in a wide range of ice conditions. After several iterations on hardware and software design with wave tank experiments confirming its capability up to 1.28 Hz, 35 buoys have since been built and used in practical and scientific applications (Pärt et al. 2023; Rikka et al. 2024). The current version allows operations for more than 14 months with primary batteries while transferring real-time wave data over the Iridium or LTE networks every 22 minutes. The technical specifications are available at www.lainepoiss.eu.

e. Spartacus

Spartacus is a comparatively small and light spar buoy, being developed to measure bubbles and other particles close to the surface, under breaking waves. Made from a standard PVC pressure pipe, it has 11 cm outside diameter and 6 m length. Bubbles and other particles are measured by imaging with the Silhouette Camera previously developed by SINTEF (Davies et al. 2017). The camera, telecentric lens, computer and batteries are housed inside the body of the buoy. The camera and lens look out of the buoy via a 45º mirror and a window, and directly onto a white, uniform LED backlight. The system captures bright-field images, where any particles show up as dark against a white background. Every image captures a known volume of water, allowing concentrations of particles and bubbles to be calculated.

Unlike the buoys discussed above, the purpose of Spartacus is not to directly measure waves, but to provide a relatively stable and minimally invasive platform, in limited wave conditions, from which other parameters can be measured at known average depth under breaking waves. In the first deployment in spring 2024, we obtained sharp images of entrained bubbles in wind up to 20 ms$^{-1}$ gusts. In addition to the described bubble and particle measurements, other sensors can be mounted at intervals along the body of the buoy to get near-surface profiles.

f. Other similar buoys in the literature

We also noted during WOOP24 that many more buoys with generally similar designs to the ones discussed at the workshop are currently under development by a variety of groups. While it is challenging to come up with an exhaustive list, we can point to, e.g., the ultra-light (total weight 52 g) MEMS-based short wind-wave sensing buoy of Yurovsky and Dulov (2020) that can be deployed from a fishing rod, the ultra-low-cost education-focused smart-buoy from T3chFlicks (2021), the FZ series of buoys (Kodaira et al. 2022, 2024), or the ultra-compact wavedrifter that has been used to measure wave plungers by Feddersen et al. (2023). These come in addition to established commercial affordable buoys, such as the already mentioned Sofar Spotter (Raghukumar et al. 2019), and new emerging buoys as MELODI (Charron and Mironov 2023). Collins et al. (2024) presents a review and inter-comparison of several small wave buoys now in common usage.

## 3. Experiments and applications

a. Tropical cyclones and extreme conditions

Tropical cyclones (e.g., hurricanes, typhoons) are notoriously difficult to sample, given their rapid evolution in time and localization in space. Existing networks of moored buoys are unlikely to capture the full complexity of these storms. Targeted deployments of mini-buoys are a promising approach to fill this gap. This is the approach of the NOPP Hurricane Coastal Impacts project https://nopphurricane.sofarocean.com, which deploys Spotters, microSWIFTs, and air-deployed wave surface drifters (ADWSD) from aircraft into the path of land-falling hurricanes. The first results from this project confirm the saturation of wave slopes (and the sea surface roughness that causes drag) at high winds (Davis et al. 2023). Ongoing work uses dense arrays of mini-buoys to explore spatial patterns within the storms, including effects of wind-wave alignment on wave slopes and drag. These results require spatially distributed and synoptic measurements of waves within the storm, and thus are an ideal use case for small wave buoys.

b. Wave-ice interactions

Many previous studies have used buoys to measure wave attenuation in sea ice (Rogers et al. 2016; Thomson et al. 2018; Cheng et al. 2017; Kohout et al. 2020). These studies had typically less than ten buoys, partially due to the high cost of full-sized wave buoys. The low number of buoys has prevented these studies from resolving strong gradients at the ice edge (Hosekova et al. 2020), forcing the assumption of homogeneity in the attenuation, something probably not warranted (Herman 2024). Specifically, the classic formulation for an exponential decay of wave energy following a spatially homogeneous attenuation rate is ripe for revision. The advent of low-cost mini-buoys that can be deployed in large numbers (>10) has the potential to transform our understanding of both wave growth and attenuation in the presence of sea ice, and to provide more representative datasets of waves in polar environments. The deployment of low cost buoys (including earlier versions of the OMB) has already had an impact on wave-ice-interaction studies, helping to validate and test wave-in-ice models (Voermans et al. 2021), leading to the development of criteria for ice breaking under the influence of waves (Voermans et al. 2020). These developments are further underway with deployments of the OMB (Rabault et al. 2023, 2024; Müller et al. 2024) and microSWIFTs (Thomson et al. 2023), and are already revealing physical mechanisms at play in the marginal ice zone (Rabault et al. 2024), which has recently resulted in improvements to the accuracy of operational models (Aouf et al. 2024). Similarly, model weaknesses have been revealed by in-situ data (Nose et al. 2023, Dreyer et al. 2024).

c. Wave-current interactions

Wave-current interactions are another topic with complex spatial gradients which will benefit from distributed sampling with large numbers of buoys. Previous work has used small numbers of drifting buoys to observe increases in wave steepness, wave breaking, and wave dissipation in the presence of strong opposing currents (Thomson et al. 2014; Zippel and Thomson 2017; Zippel et al. 2018; Iyer et al. 2022). Their experiments in the Columbia river mouth have shown clearly how a denser deployment can shed light on the impact of vertical current shear on wave-breaking dissipation. Repeating these experiments with denser arrays will undoubtedly reveal new physics, especially in regions with strong mesoscale variability (Ardhuin et al. 2017; Kudryavtsev et al. 2017). The spatial coverage provided by arrays of small buoys will advance understanding of changes that occur on spatial scales of a few wavelengths, including breaking and blocking (Chawla and Kirby, 2002). The effect of strong vertical shear is another topic that large numbers of buoys may illuminate (Banihashemi and Kirby 2019; Kirby and Chen 1989). The vertical shear sets the wavelengths which respond to the currents, which in turn sets the scale of wave breaking and subsequent release of wave momentum and energy.

d. Wave breaking and surf-zone dynamics

Wave breaking in both deep and shallow water has long been a key challenge to observe in situ. As shown by the stereo-video work of Schwendeman and Thomson (2017), buoys, especially if large, require an assumption of linear dispersion to infer the steepness of propagating waves, and this obscures the evolution of individual crest steepening that leads to breaking. This becomes more severe in the coastal surf zone, where spatial and temporal gradients are likely to be higher. Recently, Rainville et al. (2023) used arrays of 50 microSWIFT buoys deployed simultaneously in the surf zone to reconstruct sea surface elevations, detect breaking, and investigate material transport during breaking. Up to $10g$ accelerations have been measured in the surf zone (Hope et al. 2024), where large arrays of small buoys can at least identify the spatial and temporal gradients so relevant for the local dynamics and evolution. Indeed, one of the advantages of these small buoys is the possibility of adapting, hence optimizing, their performance to the environment we want to explore.

e. Entrainment by breaking waves

Breaking waves play an important role in exchange of mass and momentum between the ocean and the atmosphere, and also contribute to mixing in the upper water column. An important example is the absorption of $CO_2$ by the ocean, where bubbles created by breaking waves play an important role, while simultaneously being hard to measure in the field (Deike 2022). Similarly, during marine oil spills, breaking waves and vertical mixing have a key role in determining the fate of the spilled oil (Johansen et al. 2015; Röhrs et al. 2018). The new generation of buoys described here, by virtue of being small, cheap, and easy to deploy, can provide dense observations under a range of conditions, including during oil spills. With a buoy like Spartacus, which can also be deployed by hand from a small boat, entrainment of bubbles, oil droplets and other particles can be measured at multiple depths. When combining wave measurements with entrainment measurements, truly novel datasets can be achieved, that can push our understanding and our parameterizations of air bubble and oil droplet entrainment. A recent example is the characterization of bubble plume depths measured by echograms beneath drifting SWIFT buoys, which show strong relations of bubbles to wave conditions and wave dissipation (Derakhti et al. 2024).

## 4. Outlook

What we described above is only a hint at the possibilities offered by the new technology. But technology keeps improving, and it is reasonable to expect further miniaturization of these floating buoys in the years to come. This comes with the advantages we have illustrated, but also with a number of challenges for the possibly different hydrodynamic behaviors of very small buoys.

One of the exciting aspects of deploying miniature wave buoys is the possibility of combining them with other instruments. Traditional oceanographic experiments have tended to involve large buoys and fixed moorings. Small, expendable buoys that report back either phase-averaged spectra or full time series at high temporal resolution, can more easily be combined with instruments that measure remotely from space or from fixed locations, such as stereo cameras (Benetazzo 2006, Benetazzo et al. 2012), radar altimeters (Ewans et al. 2014), down-looking lasers (Malila et al. 2022a), and nautical radars (Malila et al. 2022b).

Buoys have long been used for measurements in the surf zone. As shown in sub-section 3d, size matters. It is therefore important to make programs in this field of research. The smaller the better, and buoys should ideally be "sticky", i.e., water-following, or rather surface-following. There is progress in this field, with Feddersen et al. (2023) presenting a tennis-ball sized buoy (but without the telemetry required to communicate the data back to shore) and, as mentioned, the SFY (Hope et al. 2024) measuring acceleration in surf-zone breakers in excess of $10g$, something made possible by being at the right spot at the right time. It seems clear that shape, density and center of gravity will remain important factors as we continue to shrink our equipment, until we get to millimeter scale where turbulence (e.g. breakers) may dominate over floating capability.

Arrays (organized) or swarms (unorganized) of drifting buoys show great promise in mapping wave-current interaction in regions with strong current gradients. Indeed, in regions with very strong currents, e.g. the Moskenes current in northern Norway, one of the strongest open-ocean tidal currents in the world (Gjevik et al. 1997; Saetra et al. 2021; Halsne et al. 2022), the only feasible way of simultaneously mapping the flow and the wave field is by deploying a large number of (potentially phase-resolving) drifting buoys. These processes are still poorly understood, and the related study is still only in its infancy (Halsne et al. 2023).

The importance of expendability becomes crucial when considering experiments in the sea ice. Here, buoys have been considered expendable for a long time. Reducing the cost by an order of magnitude, deployments of OMB arrays have demonstrated that wholly new types of experiments can be conducted with large numbers deployed over periods of weeks or months (Rabault et al. 2022).

The recent proliferation of unmanned aerial systems, or drones, opens up the possibility of easily deploying lightweight buoys from shore or from research vessels (see, e.g., Alari et al. 2022). Drones also offer an appealing vantage point for remote sensing which dove-tails with surface buoys. We expect drones and miniature buoys to become a scientific power couple in the years to come.

Looking to the future, what can we expect ten years from now? A reasonable guess is that what is now a 10-15 cm box will be more or less the size of a table tennis ball (see Feddersen et al. 2023 for a first example of a wave buoy approaching such a degree of miniaturization, albeit, as mentioned, without telemetry). Such buoys will also combine the complementary strengths of both GNSS and IMU-based measurements by using increased on-board computational power to perform full data fusion and reach even higher accuracy across the whole wave frequency spectrum. At this point the first major advantage is the increased expendability (neglecting the negative aspect of pollution). Large sets of these buoys can be deployed in a limited area providing a detailed three-dimensional and time view of the surface, starting from the initial conditions and with known wind characteristics. The drawback of these large expendable sets is that they will soon drift away from each other. Possibly something to work on is a system of communication and propulsion that would "pull" the buoys towards each other, or align them on a grid, counteracting the randomness of the sea pushing for further distances. Calibration of the distance would lead to different experiments and analyses of the situation, at different scales and resolution.

While oceanographic parameters (waves, current, temperature, etc) are obvious possibilities, a key question concerns the wind. A small companion spar buoy could provide both the wind close to the sea surface and currents in the upper layers. However, what we are really after is the stress at the surface. That is the true connection and what is felt by both the atmosphere and the ocean. The miniaturization leads us closer and closer to the separation surface, where the true coupling acts.

A key aspect of these modern buoys is the focus of several of the designs on open source software and hardware. This helps cutting the cost per buoy and improves scientific reproducibility and reliability, since anybody can inspect and improve on the design. This also provides a reference platform to easily perform measurements of additional physical parameters, reducing the barrier to entry for new research groups. This also strongly reduces the barrier to entry for private companies to produce their own low cost buoys, benefitting from all the complex, time consuming, and costly research and development provided freely by the scientific community. This will likely result in a drastic increase in competition and a corresponding drop in price also of commercial instrumentation in the coming years.

While the availability of more and higher resolution data is interesting per se for fundamental oceanographic studies, it also increasingly enables new methods from the realms of statistics, machine learning and data-driven methods to be applied to old problems. The recent "Forecasting Floats in Turbulence Challenge" organized by the Defense Advanced Research Projects Agency (DARPA, see https://www.darpa.mil/news-events/2021-12-13), based on trajectories of Sofar Spotter buoys, is a case in point. With increasing numbers of low cost buoys being deployed, we expect more opportunities will arise in the future.

We end with a challenge from one of the greats of physics. In Richard Feynman's 1959 lecture, "There's Plenty of Room At The Bottom" (Feynman 1961), he laid out the enormous potential for future discoveries that would come with miniaturization. So, at our more mundane dimensions seven or eight orders of magnitude above the nanoscale envisioned by Feymann, let's go forth and miniaturize our buoys! There is plenty of "room at the bottom", for us at the top of the ocean, and we can rest assured that Munk was right in his intuition that new discoveries will follow new technologies, especially so if we keep our science, our software, and our hardware open source.

**Acknowledgments**

This work and the WOOP24 workshop at Finse was supported by the Research Council of Norway through the ENTIRE project, grant no 324227. The authors also gratefully acknowledge additional funding from the Norwegian Meteorological Institute for the organization of the workshop. Additional travel support provided by the Valle Exchange program at the University of Washington.

**Data availability**

All data are already published in the references cited – no new data are presented.

**Acronyms**

| | |
|---|---|
| ADWSD | air-deployed wave surface drifters |
| DARPA | defence advanced research project agency |
| FOSS | free open source software |
| FOSSH | free open source software and hardware |
| GNSS | global navigation satellite system |
| IMU | inertial measurement unit |
| LED | light emitting diode |
| LTE | long-term evolution (transmission) |
| MCU | micro-controller unit |
| OMB | open met buoy |
| SFY | small friendly buoy |

SWIFT        surface wave instrument float with tracking

TCO        total cost of ownership